# Report on the 2019 Workshop on Smart Farming and Data Analytics (SFDAI)

**Liadh Kelly**
Computer Science Department
Maynooth University
Ireland
*liadh.kelly@mu.ie*

**Simone van der Burg**
Wageningen Economic Research
The Netherlands
*simone.vanderburg@wur.nl*

**Aine Regan**
Teagasc
Ireland
*Aine.Regan@teagasc.ie*

**Peter Mooney**
Computer Science Department
Maynooth University
Ireland
*peter.mooney@mu.ie*

**Abstract**

The 1st National workshop on Smart Farming and Data Analytics took place at Maynooth University in Ireland on June 12, 2019. The workshop included two invited keynote presentations, invited talks and breakout group discussions. The workshop attracted in the order of 50 participants, consisting of a mixture of computer scientists, general scientists, farmers, farm advisors, and agricultural business representatives. This allowed for lively discussion and cross-fertilization of ideas. And showed the significant interest in the smart farming domain, the many research challenges faced in the space and the potential for data analytics and information retrieval here.

## 1 Introduction

Smart Farming represents the application of modern Information and Communication Technologies (ICT) into agriculture. Smart Farming sees the combined application of ICT solutions such as Information Retrieval, precision equipment, the Internet of Things (IoT), sensors and actuators, geo-positioning systems, Big Data, Unmanned Aerial Vehicles (UAVs, drones), robotics, etc. Practically, Smart Farming should provide the farmer with added value in the form of better decision making, new knowledge about their farm or more efficient operations and management of the farm itself.

Globally there is a growing awareness of Smart Farming. However, the understanding and focus of most 'Smart Farming' revolves around more efficient operations and management of the farm itself. There is much less emphasis on using data and information generated by the farm (machinery, collection of data from farm animals, farm sensors, etc.) to support farmers, informing better decision making and new knowledge. In 2016 the OCED reported that 80% of data collected on farms is never "stored or accessed in a way that the data becomes actionable".

This is where the 1st National Workshop on Smart Farming and Data Analytics in Ireland (SFDAI) comes into focus. The two primary goals of the workshop were to:

(1) highlight to farmers and the agricultural community the innovations possible in Ireland through an exploration of data analytics for Smart Farming, and
(2) to interact with and learn from the farming community so that researchers and scientist can develop a research agenda for Smart Farming data analytics in order to bring about the practical reality of Smart Farming and Data Analytics.

While this workshop was tuned to an Irish audience, its message and outcomes are applicable to the global Smart Farming community. Of particular interest are the potentials for the information retrieval research community to contribute to this domain through exploitation of data being gathered on farms in intelligent ways – for example in intelligent search, decision making and other types of support agents.

## 2   Keynote 1 – Thia Hennessy

The first keynote talk was given by Prof Thia Hennessy. Prof Hennessy is Dean of the Business School at University College Cork (UCC), Ireland. She also holds a number of directorships, serving on the boards of Teagasc, the Agriculture and Food Development Authority of Ireland, the Irish Management Institute and the Cork Chamber of Commerce. She is also Professor and Chair of Agri-Food Economics in UCC. Over her research career she has examined issues such as the impact of changing agricultural policy on farm performance, the implications of environmental policy for agriculture and food production and the role of technological developments in farm performance.

In this keynote the global imbalance in food was highlighted, where people are going hungry when others have too much and there are large amounts of waste; there is global warming; and poor migrant workers reliance on volatile UK and USA markets. Technology in general can help with these issues: (1) what we eat (e.g. lab grown burgers); (2) how it is produced (e.g. drones); (3) where it is produced (e.g. vertical farms); (4) who produces it – there is a shrinking role for humans here, instead have autonomous vehicles, robotic milkers and fruit pickers for example; (5) how it is consumed (e.g. eating supplements instead of food). There is need for the integration of many technologies and progress already being made, e.g. 3D printing of food, machinery and parts, blockchain, augmented reality (AR) and virtual reality (VR), robotics, artificial intelligence (AI), etc. Overall there is huge opportunity for the use of data analytics in smart farming solutions and ability to generate more data, to inform needed better decision making.

## 3   Keynote 2 – Darragh McCullough

The second keynote talk was given by Mr Darragh Mc Cullough, a well know figure in the Irish farming community who is himself both an active farmer and regular contributor to Irish farming media both written and televised. He presents the popular Ear to the Ground[1] programme on Irish television and writes a weekly column for an Irish newspaper's farming section[2].

---

[1] https://www.rte.ie/tv/programmes/913566-ear-to-the-ground/
[2] https://www.independent.ie/business/farming/

Darragh highlighted the existing use of technology on farms to support such things as robotic dairy farming, machines to bunch flowers and pump the right type of air into glass houses, autonomous solar powered weeders, camera powered hoes, sensors on machines that exploit AI, etc. Such technology are economical and allow for less waste. While technology is already impacting farming, there is much scope for further developments. The main driving forces for such technology are: (1) environment; (2) food safety; (3) welfare (of animals, people) (4) labor; (5) cost. Challenges are: (1) does it pay?; (2) is it useable (can machines interact with other machines, can they interact with data, can the farmer interact with data); (3) will technology drive a bigger wedge between the haves and have nots.

# 4 Invited Talks
The workshop featured three invited talks.

## 4.1 What can Drones do?

The first invited talk was given by Mr Aidan Magee (U-Flyte Project, National Center for Geocomputation, Maynooth University, Ireland) on 'What can Drones do?'.
Within recent years drones, have seen rapid growth in popularity across a variety of industries including the agricultural. This talk discussed how drones can be used as a tool to enable precision agriculture. The talk also explored the practical and regulatory considerations needed when using drones, and examined the viability of their use within Ireland today.

## 4.2 How to make Sense of your Data

The second invited talk was given by Dr Christina O'Connor (School of Business, Maynooth University, Ireland) on 'How to make Sense of your Data'. Christina has worked with farm businesses across the UK and Ireland over the last 10 years.
This talk discussed and explored the following areas: (1) What is Big Data? (2) What is the value of Big Data? (3) What is the process in making sense of Big Data? The volume and variety of data deriving from everyday activity on the farm is an excellent source of insight to the direction of the farm, but this is largely based on making sense of this data and understanding where value can be added.

## 4.3 Exploring Copernicus

The third invited talk was given by Dr Conor Cahalane (Department of Geography, Maynooth University, Ireland) on 'Exploring Copernicus'.
Copernicus is the European Union's satellite mapping programme. This talk showed the benefits that satellites can bring to Agriculture - helping when assessing agricultural land use and trends, crop conditions and yield forecasts. Copernicus supports irrigation management, seasonal mapping of cultivated areas, water management and drought monitoring, as well as subsidy controls.

# 5 Discussion Session
Three breakout groups discussed the topic of data ethics associated with smart farming, in breakout groups organized by Simone van der Burg, Wageningen Economic Research, The Netherlands and Aine Reegan, Teagasc, Ireland. A fourth breakout group discussed the potential for smart farming data analytics on small to medium size Irish farms.

## 5.1 Data Ethics

Half of the workshop attendees were split into three separate groups to discuss, in a structured way, data ethics associated with smart farming. Each group was presented with four different storylines for future scenarios for data governance and asked to consider their pros and cons and indicate their own preference. Scenarios were:

- The 'I choose model' in which a farmer should get to choose whether and with whom he or she shares data.
- 'Data as public library' where management of the 'library'/hub develops sharing policy.
- 'Data governance is settled by the market' where farmers and other businesses will share data when it brings them benefits.
- 'Data allow to re-organize collaboration in the value chain' where farmers shape collaboration with supply-chain and supermarket.

Much interest was shown in storyline 2 (the digital library) and 1 (the 'I choose model'). The 'I choose model' was considered interesting as it allows farmers to choose for themselves, which seems appropriate as it is their data. Participants felt it would be strange not to ask their consent. But the limitations of consent were also acknowledged, in a world with an increasingly large data sharing network. In such a world, consent likely becomes a formality without much content, much like signing a privacy agreement while downloading an app on the phone. Some participants suggested substituting the informed consent agreement by something else; like, a license that companies can obtain when they respect certain norms. This helps farmers to consent, as the license tells them the company will respect these norms and they will not have to read information about what the company is going to do with the data, which may be complex and voluminous. For companies it is also clear, as they can apply for the license and therewith enhance trust in their business.

There was also interest in the 'digital library'. Some preferred this model as this helps to keep data available for public actors, such as the government or research. Others thought that companies should also go to digital libraries to get their data. They preferred the digital libraries as they thought farmers would then only have to give their consent once: they would have to learn about the data sharing policy of the library, give their consent to that, and then the library would govern the data after that. In this way, farmers would not have to give consent to every individual actor asking for their data, having to invest time and energy into reading and evaluating the information that this actor provides. This task is delegated to the digital library.

Some participants, however, objected that it would be hard to give the data directly to the library, as they are often generated by machinery provided by the companies. So the companies generate data and afterwards data will have to be moved to the library. Will companies be willing to do that? Some people doubted that and said that, actually, farmers decide to hand over the ownership of their data to the company when they start using IOT at their farms. There's no way around that. Others suggested having an intermediary between the digital libraries, the companies and the farmers. This intermediary would have to mediate between the interests of the tech companies, the farm businesses and the public goals served with digital farming. They would have to decide what data should go to the digital library for secondary use by researchers and policy makers. Some objected, however, that companies do not have an interest to hand over data to a library. This would have to be a government's decision. Or it could be regulated with some kind of trademarks or licenses: if you decide to make your data available to a digital library, than you show your support to public goals, and you deserve a trademark or license. In this way, companies could become interested in sharing data, as it gives them a positive image.

## 5.2    Potential for Smart Farming Data Analytics

This discussion group was split into 4 groups for breakout group discussion. Following 40 minutes of breakout group discussion, participants regrouped to report back and conduct large group discussion (duration 60 minutes). Each breakout group discussed how smart farming data analytics might help small to medium size Irish farmers on their farms (dairy, tillage, suckler, etc.) under one of the following 4 topics:

- Environment and climate
- Safe farming
- Economic return
- Social and cultural

Discussion points considered for these topics were:

- What do farmers most need support with?
- What challenges might smart farming data analytics be able to address?

**Outcomes:**
Some key outcomes from this discussion were the fact that to maximize economic return farmers need more return from their assets. Farmers will get the most value from lean and smart farming and from timeliness and speed. Safety on farms is of paramount importance. Useful from a technology point of view would be real time information for farm dangers. For example, sensors in sheds telling a farmer what is going on when they are doing slurry.

Data analytics and smart farming might contribute solutions for alleviating pressure on farmers. Grass management is of environmental interest and still a primary issue on Irish farms and everything else stems from this. Farmers want to know when to graze, cut, etc. Smart farming sensors can analyse soil and water.  New developments in data analytics have the potential to produce maximum output for the farmer with minimum input, which can contribute to alleviating the decision making pressures faced by farmers.

While many smart technologies already exist, a key challenge is how to get data talking to each other. Who is delivering the technology solution message is also important for farmers to trust the technology.  Farmers are also fearful that they will lose the ability to farm because of technology. That is they will lose the knowledge of how to relate to their land and animals, for example the ability to look at grass and know it is ready to cut. However, there is a counter argument that both farmer and technology can work in harmony. Indeed this should be a core goal of any smart farming developments. They should aim to supplement and support the farmer, as opposed to replacing his or her skills. A lot of small farms are engaged with farming for the love of farming, so they do not want their farming practices such as looking at their fields to be replaced by technology. Technology should complement and support the farmer's knowledge, and not replace it.

# 6  Conclusions

This was the first national workshop on Smart Farming and Data Analytics held in Ireland. While the workshop largely consisted of an Irish audience, with discussion tuned to Smart Farming in Ireland, the take home messages from the workshop have global reach. The volume of interest in the workshop, highlight the interest in this topic not only within the computer science community, in particular the information retrieval and data analytics communities, but also across scientists, farmers, farm advisors and agricultural business representatives. Given the wide breath of participation in the workshop it provided a unique opportunity for computer scientists to gain insights into the pressing requirements of the farming community and opportunity to exploit new information retrieval and data analytics techniques to offer smart farming solutions on farms.

# 7  Acknowledgements

This workshop was supported by: Department of Computer Science, Maynooth University; Department of Geography, Maynooth University; the U-Flyte project and National Center for Geocomputation, Maynooth University; the School of Business, Maynooth University; The National Rural Network Ireland, and Society of Chartered Surveyors Ireland (SCSI). Thanks also go to the keynote speakers, invited speakers and workshop attendees, without whom the workshop would not have been the success it was.